\def\be{\begin{equation}}
\def\ee{\end{equation}}
\def\ba{\begin{array}}

\def\ea{\end{array}}

\documentclass[12pt]{article}
\usepackage{amsfonts,amssymb,amsmath,graphicx}
\begin{document}
\parskip=3pt
\parindent=18pt
\baselineskip=20pt
\setcounter{page}{1}
\centerline{\large\bf Total variance and invariant information}
\centerline{\large\bf in complementary measurements}
\vspace{6ex}
\centerline{{\sf Bin Chen,$^{\star}$}
\footnote{\sf Corresponding author: chenbin5134@163.com}
~~~ {\sf Shao-Ming Fei$^{\natural,\sharp}$}
}
\vspace{4ex}
\centerline
{\it $^\star$ College of Mathematical Science, Tianjin Normal University, Tianjin 300387, China}\par
\centerline
{\it $^\natural$ School of Mathematical Sciences, Capital Normal University, Beijing 100048, China}\par
\centerline
{\it $^\sharp$ Max-Planck-Institute for Mathematics in the Sciences, 04103 Leipzig, Germany}\par
\vspace{6.5ex}
\parindent=18pt
\parskip=5pt
\begin{center}
\begin{minipage}{5in}
\vspace{3ex}
\centerline{\large Abstract}
\vspace{4ex}

We investigate the total variance of a quantum state with respect to a complete set of mutually complementary measurements and its relation to the Brukner-Zeilinger invariant information.
By summing the variances over any complete set of mutually unbiased measurements and general symmetric informationally complete measurements respectively,
we show that the Brukner-Zeilinger invariant information associated with such types of quantum measurements is equal to the difference between the maximal variance and the total variance obtained. These results provide an operational link between the previous interpretations of the Brukner-Zeilinger invariant information.
\end{minipage}
\end{center}

\newpage

\section{Introduction}

In a classical measurement, the observation
removes our ignorance about the state by revealing the
properties of the system which are considered to be pre-existing
and independent of the observation. The Shannon information is a desirable measure to quantify the amount of information carried by a classical system. It is also
a natural measure of our ignorance regarding the properties of a classical system.
However, the situation is quite different for the case of quantum measurements, for which
one cannot say that quantum measurements reveal a pre-existing property of a quantum system.
Therefore, the Shannon entropy could be thought of as ¡°conceptually¡± inadequate
in quantum physics. Brukner and Zeilinger introduced a quantity as a new measure of total information obtained by summing individual measures over a complete
set of mutually complementary measurements \cite{BZ1,BZ2}.
This measure of quantum information takes into account that
the only features of quantum systems known before a measurement
are the probabilities for various events to occur.
This quantity can be expressed as
\begin{equation}
I(\rho):=\mathrm{Tr}(\rho^{2})-\frac{1}{d},
\end{equation}
where $d$ is the dimension of the quantum system.

Moreover, the ``total" uncertainty related a quantum measurement is also defined.
For $d$ measurement outcomes with probabilities $p_1,p_2,...,p_d$,
the lack of information about the $j$-th outcome with respect to a single experimental trial is given by $p_j(1-p_j)$. The total lack of information regarding all $d$ possible experimental outcomes is then given by $\sum_{j=1}^{d} p_j(1-p_j)=1-\sum_{j=1}^{d} p_j^2$, which is nothing but $1-\mathrm{Tr}(\rho^{2})$, where $\rho$ is the state after a quantum
(projective) measurement, the linear entropy of the measured state.
The ``total" uncertainty related to a quantum measurement is then given by
\begin{equation}
U(\rho):=1-\mathrm{Tr}(\rho^{2}),
\end{equation}
which is the sum of all individual measurement uncertainties over a complete set of mutually complementary measurements.

It can be seen that $I(\rho)$ and $U(\rho)$ are two mutually complementary quantities, and both of them are invariant under unitary transformations.
These two quantities have many useful applications in various issues of quantum mechanics and quantum information theory \cite{1,2,3}.
In the rest of this paper we call $I(\rho)$ the BZ invariant information and $U(\rho)$ the BZ invariant uncertainty.

In Ref. \cite{Luo}, Luo presented an alternative interpretation of the BZ invariant information.
By summing the variances of a quantum state over any complete orthogonal set of observables, he obtained a new quantity
\begin{equation}
V(\rho):=d-\mathrm{Tr}(\rho^{2}).
\end{equation}
This quantity, called total variance, is related to the BZ invariant information and the BZ invariant uncertainty.
One can see that $V(\rho)$ attains its minimum value $V_{\mathrm{min}}(\rho):=d-1$ iff $\rho$ is a pure state,
and attains its maximum value $V_{\mathrm{max}}(\rho):=d-1/d$ iff $\rho=I/d$ is the maximally mixed state.
Hence the BZ invariant information is equal to the difference between the maximal variance and the total variance,
while the BZ invariant uncertainty is exactly the difference between the total variance and the minimal variance, i.e.,
\begin{equation}
I(\rho)=V_{\mathrm{max}}(\rho)-V(\rho),~~~U(\rho)=V(\rho)-V_{\mathrm{min}}(\rho).
\end{equation}
To this point, the BZ invariant information (uncertainty) can be viewed as a renormalized version of the total variance of a quantum state \cite{Luo}.

After that, Rastegin reinterpreted the BZ invariant information from another perspective \cite{Ras}.
By introducing the index of coincidence \cite{co}
\begin{equation}
C(\mathcal{P}|\rho):=\sum_{j}p_{j}(\mathcal{P}|\rho)^{2},
\end{equation}
the BZ information measure can be rewritten as
\begin{equation}
I(\mathcal{P}|\rho)=C(\mathcal{P}|\rho)-C(\mathcal{P}|\rho_{\ast}),
\end{equation}
where $\mathcal{P}$ denotes a set of complementary measurements,
$\{p_{j}(\mathcal{P}|\rho)\}_{j}$ is the probability distribution given by $\mathcal{P}$ on the quantum state $\rho$,
and $\rho_{\ast}$ is the maximally mixed state.
When a complete set of $d+1$ mutually unbiased bases (MUBs) \cite{WF,Durt} in $d$-dimensional quantum system is taken into account, one gets \cite{La,Iv}
\begin{equation}
C(\mathcal{P}|\rho)=1+\mathrm{Tr}(\rho^{2}),
\end{equation}
and in this case
\begin{equation}
I(\mathcal{P}|\rho)=\mathrm{Tr}(\rho^{2})-\mathrm{Tr}(\rho_{\ast}^{2})=\mathrm{Tr}(\rho^{2})-\frac{1}{d},
\end{equation}
which is in consistent with the original definition of the BZ invariant information.

Besides MUBs, there are several special types of quantum measurements,
the mutually unbiased measurements (MUMs, as a generalization of MUBs) \cite{KG1},
the symmetric informationally complete positive operator-valued measures (SIC-POVMs) \cite{Re},
and the general symmetric informationally complete measurements (general SIC measurements, as a generalization of SIC-POVMs) \cite{Ap,GK2}.
Both of the MUMs and the general SIC measurements form a complete set of mutually complementary measurements respectively.
It has been shown that for a complete set of $d+1$ MUMs $\mathcal{P}$ with the parameter $\kappa$, the BZ information measure gives rise to \cite{Ras}
\begin{equation}
I(\mathcal{P}|\rho)=\frac{\kappa d-1}{d-1}[\mathrm{Tr}(\rho^{2})-\mathrm{Tr}(\rho_{\ast}^{2})],
\end{equation}
while for a complete set of $d^{2}$ general SIC measurements $\mathcal{M}$ with the parameter $a$, one has \cite{Ras},
\begin{equation}
I(\mathcal{M}|\rho)=\frac{ad^{3}-1}{d(d^{2}-1)}[\mathrm{Tr}(\rho^{2})-\mathrm{Tr}(\rho_{\ast}^{2})].
\end{equation}

Then a natural question is what the relationship between these two interpretations of the BZ invariant information is.
Inspired by Luo's work, in this paper, we define the notion of total variance in a set of complementary measurements.
By summing over all variances of a quantum state for a complete set of MUMs and general SIC measurements, respectively,
we show that the resulted total variance is related to BZ invariant information for such types of quantum measurements given in \cite{Ras}.
In particular, such BZ invariant information is precisely the difference between the maximal variance (total variance of the maximally mixed state) and the total variance.
Our results provide an operational link between the two interpretations of BZ invariant information.

\section{BZ invariant information in MUMs}

Let us first review some basic concepts about MUBs and MUMs.
Two orthonormal bases in $d$-dimensional Hilbert space $\mathcal{H}_{d}$ are said to be mutually unbiased if the absolute values of the inner products of any vector from one basis and any vector from another basis are $1/\sqrt{d}$.
A set of orthonormal bases in $\mathcal{H}_{d}$ is called a set of MUBs if each two bases from the set are mutually unbiased.
It has been shown that the maximum number $N(d)$ of MUBs is no more than $d+1$, and $N(d)=d+1$ when $d$ is a prime power \cite{WF}.
But $N(d)$ is still unknown when $d$ is not a prime power \cite{Durt}.
In Ref. \cite{KG1}, Kalev and Gour generalize the concept of MUBs to MUMs.
They show that there always exists a complete set of $d+1$ MUMs for arbitrary $d$, which can be explicitly constructed.
Two POVM  measurements $\mathcal{P}^{(b)}=\{P_{n}^{(b)}\}_{n=1}^{d}$, $b=1,2$, are said to be MUMs if
\begin{equation}
\begin{split}
\mathrm{Tr}(P_{n}^{(b)})&=1,\\
\mathrm{Tr}(P_{n}^{(b)}P_{n'}^{(b')})&=\frac{1}{d},~~~b\neq b',\\
\mathrm{Tr}(P_{n}^{(b)}P_{n'}^{(b)})&=\delta_{n,n'}\,\kappa+(1-\delta_{n,n'})\frac{1-\kappa}{d-1},
\end{split}
\end{equation}
where $\frac{1}{d}<\kappa\leq1$, and $\kappa=1$ if and only if all $P_{n}^{(b)}$'s are rank one projectors, which gives rise to two MUBs.
The construction of a complete set of $d+1$ MUMs is as follows.
Let $\{F_{n,b}:n=1,2,\ldots,d-1,b=1,2,\ldots,d+1\}$ be a set of $d^{2}-1$ Hermitian, traceless operators acting on $\mathcal{H}_{d}$, satisfying $\mathrm{Tr}(F_{n,b}F_{n',b'})=\delta_{n,n'}\delta_{b,b'}$.
Define $d(d+1)$ operators
\begin{equation}\label{2}
F_{n}^{(b)}=
\begin{cases}
   F^{(b)}-(d+\sqrt{d})F_{n,b},&n=1,2,\ldots,d-1;\\[2mm]
   (1+\sqrt{d})F^{(b)},&n=d,
\end{cases}
\end{equation}
where $F^{(b)}=\sum_{n=1}^{d-1}F_{n,b}$, $b=1,2,\ldots,d+1$.
Then the operators
\begin{equation}\label{3}
P_{n}^{(b)}=\frac{1}{d}I+tF_{n}^{(b)},
\end{equation}
with $b=1,2,\cdots,d+1,n=1,2,\cdots,d$, give rise to $d+1$ MUMs, where $t$ should be chosen such that $P_{n}^{(b)}\geq0$.
Thus the parameter $\kappa$ is given by
\begin{equation}\label{5}
\kappa=\frac{1}{d}+t^{2}(1+\sqrt{d})^{2}(d-1)
\end{equation}
from the construction.
On the other hand, any $d+1$ MUMs can be expressed in such form \cite{KG1}.

A complete set of $d+1$ MUMs is exactly a complete set of mutually complementary measurements,
and can be used to characterize the BZ invariant information in measurements.
To this end, we define the quantity $V(\mathcal{P}_{\mathrm{MUM}}|\rho)$ as a measure of total variance of a quantum state $\rho$ in a complete set of MUMs
$\mathcal{P}_{\mathrm{MUM}}=\{\mathcal{P}^{(b)}\}_{b=1}^{d+1}$ with the parameter $\kappa$,
\begin{equation}
V(\mathcal{P}_{\mathrm{MUM}}|\rho):=\sum_{b=1}^{d+1}V(\mathcal{P}^{(b)}|\rho)=\sum_{b=1}^{d+1}\sum_{n=1}^{d}V(P_{n}^{(b)}|\rho),
\end{equation}
where $V(X|\rho)=\langle X^{2}\rangle_{\rho}-\langle X\rangle_{\rho}^{2}$ is the variance of the observable $X$.
In the next, we calculate the quantity $V(\mathcal{P}_{\mathrm{MUM}}|\rho)$.

We first notice that
\begin{equation}
\begin{split}
\sum_{b=1}^{d+1}\sum_{n=1}^{d}V(P_{n}^{(b)}|\rho) & =  \sum_{b=1}^{d+1}\sum_{n=1}^{d}\langle(P_{n}^{b})^{2}\rangle_{\rho}-\sum_{b=1}^{d+1}\sum_{n=1}^{d}\langle P_{n}^{b}\rangle_{\rho}^{2}\\
& :=  \sum_{b=1}^{d+1}\sum_{n=1}^{d}\langle(P_{n}^{b})^{2}\rangle_{\rho}-C(\kappa,\rho),
\end{split}
\end{equation}
where $C(\kappa,\rho)$ denotes the index of coincidence for probability distribution generated by $\{P_{n}^{(b)}\}$.
It has been shown that \cite{Chen},
\begin{equation}\label{c}
C(\kappa,\rho)=\frac{(\kappa d-1)[d\mathrm{Tr}(\rho^{2})-1]+d^{2}-1}{d(d-1)}.
\end{equation}
Thus we only need to compute $\sum_{b,n}\langle(P_{n}^{b})^{2}\rangle_{\rho}$.
Taking into account that $\sum_{n=1}^{d}F_{n}^{(b)}=0,~\forall b$, we have
\begin{equation}\label{p}
\begin{split}
\sum_{b=1}^{d+1}\sum_{n=1}^{d}\langle(P_{n}^{b})^{2}\rangle_{\rho} & =  \sum_{b=1}^{d+1}\sum_{n=1}^{d}\left\langle
\left(\frac{1}{d}I+tF_{n}^{(b)}\right)^{2}\right\rangle_{\rho}\\[2mm]
& =  \frac{d+1}{d}+\frac{2t}{d}\sum_{b=1}^{d+1}\sum_{n=1}^{d}\langle F_{n}^{(b)}\rangle_{\rho}+t^{2}\sum_{b=1}^{d+1}\sum_{n=1}^{d}\langle (F_{n}^{(b)})^{2}\rangle_{\rho}\\
& =  \frac{d+1}{d}+t^{2}\sum_{b=1}^{d+1}\sum_{n=1}^{d}\langle (F_{n}^{(b)})^{2}\rangle_{\rho}.
\end{split}
\end{equation}

On the other hand,
\begin{equation}\label{f}
\begin{split}
\sum_{b=1}^{d+1}\sum_{n=1}^{d}\langle (F_{n}^{(b)})^{2}\rangle_{\rho} & =
\sum_{b=1}^{d+1}\sum_{n=1}^{d-1}\left\langle\left[F^{(b)}-(d+\sqrt{d})F_{n,b}\right]^{2}\right\rangle_{\rho}+(1+\sqrt{d})^{2}\sum_{b=1}^{d+1}\langle (F^{(b)})^{2}\rangle_{\rho}\\
& =  \sum_{b=1}^{d+1}\sum_{n=1}^{d-1}[\langle (F^{(b)})^{2}\rangle_{\rho}-(d+\sqrt{d})\langle F^{(b)}F_{n,b}+F_{n,b}F^{(b)}\rangle_{\rho}\\
&\quad  +(d+\sqrt{d})^{2}\langle F_{n,b}^{2}\rangle_{\rho}]+(1+\sqrt{d})^{2}\sum_{b=1}^{d+1}\langle (F^{(b)})^{2}\rangle_{\rho}\\
& =  (d-1)\sum_{b=1}^{d+1}\langle (F^{(b)})^{2}\rangle_{\rho}-2(d+\sqrt{d})\sum_{b=1}^{d+1}\langle (F^{(b)})^{2}\rangle_{\rho}\\
&\quad  +(d+\sqrt{d})^{2}\left\langle\sum_{b=1}^{d+1}\sum_{n=1}^{d-1}F_{n,b}^{2}\right\rangle_{\rho}+(1+\sqrt{d})^{2}\sum_{b=1}^{d+1}\langle (F^{(b)})^{2}\rangle_{\rho}\\
& =  (d+\sqrt{d})^{2}\left\langle\sum_{b=1}^{d+1}\sum_{n=1}^{d-1}F_{n,b}^{2}\right\rangle_{\rho}\\
& =  (d+\sqrt{d})^{2}(d-\frac{1}{d}).
\end{split}
\end{equation}
In the last equality, we have used the fact that $\sum_{b=1}^{d+1}\sum_{n=1}^{d-1}F_{n,b}^{2}=(d-\frac{1}{d})I$ \cite{Luo}.
Combining Eqs. (\ref{c}), (\ref{p}) and (\ref{f}), we get
\begin{equation}
V(\mathcal{P}_{\mathrm{MUM}}|\rho)=\frac{\kappa d-1}{d-1}(d-\mathrm{Tr}(\rho^{2})).
\end{equation}

It can be seen that $V(\mathcal{P}_{\mathrm{MUM}}|\rho)$ reaches its minimum value $V_{\mathrm{min}}(\mathcal{P}_{\mathrm{MUM}}|\rho)=\kappa d-1$ when $\rho$ is a pure state, and reaches its maximum value $V_{\mathrm{max}}(\mathcal{P}_{\mathrm{MUM}}|\rho)=(\kappa d-1)(d+1)/d$ when $\rho=I/d$ is the maximally mixed state.
Following Luo's perspective in Ref. \cite{Luo}, the BZ invariant information in MUMs can be characterized by
\begin{equation}
I(\mathcal{P}_{\mathrm{MUM}}|\rho):=V_{\mathrm{max}}(\mathcal{P}_{\mathrm{MUM}}|\rho)-V(\mathcal{P}_{\mathrm{MUM}}|\rho)=
\frac{\kappa d-1}{d-1}[\mathrm{Tr}(\rho^{2})-\frac{1}{d}].
\end{equation}
which is coincide with the illustration of the BZ invariant information in MUMs given in \cite{Ras}.
Furthermore, the BZ invariant uncertainty in MUMs is formulated as
\begin{equation}
U(\mathcal{P}_{\mathrm{MUM}}|\rho):=V(\mathcal{P}_{\mathrm{MUM}}|\rho)-V_{\mathrm{min}}(\mathcal{P}_{\mathrm{MUM}}|\rho)=
\frac{\kappa d-1}{d-1}[1-\mathrm{Tr}(\rho^{2})].
\end{equation}

\section{BZ invariant information in general SIC measurements}

A set of  $d^{2}$ rank one operators acting on $\mathcal{H}_{d}$ is called a SIC-POVM, if every operator is of the form
\begin{equation}
P_{j}=\frac{1}{d}|\phi_{j}\rangle\langle\phi_{j}|,~~~j=1,2,\ldots,d^{2},
\end{equation}
where the vectors $|\phi_{j}\rangle$ satisfy the following relation
\begin{equation}
\mid\langle\phi_{j}|\phi_{k}\rangle\mid^{2}=\frac{1}{d+1},~~~j\neq k.
\end{equation}
Similar to MUBs, the existence of SIC-POVMs in arbitrary dimension $d$ is still unknown.
It has been only proved that there exist sets of SIC-POVMs for all dimensions up to 67 \cite{Sco}.
In Ref. \cite{GK2}, Gour and Kalev generalize the concept of SIC-POVMs to general SIC measurements.
A set of $d^{2}$ positive-semidefinite operators $\{P_{\alpha}\}_{\alpha=1}^{d^{2}}$ on $\mathcal{H}_{d}$ is said to be a general SIC measurements, if \\
\begin{equation}
\begin{split}
\sum_{\alpha=1}^{d^{2}}P_{\alpha}&=I,\\[1mm]
\mathrm{Tr}(P_{\alpha}^{2})&=a,~\forall\alpha\in\{1,2,\ldots,d^{2}\},\\
\mathrm{Tr}(P_{\alpha}P_{\beta})&=\frac{1-ad}{d(d^{2}-1)},~\forall\alpha,\beta\in\{1,2,\ldots,d^{2}\},~\alpha\neq\beta,
\end{split}
\end{equation}
where the parameter $a$ satisfies $\frac{1}{d^{3}}<a\leq\frac{1}{d^{2}}$,
and $a={1}/{d^{2}}$ if and only if all $P_{\alpha}$ are rank one operators, which gives rise to a SIC-POVM.

Like the MUMs, there always exists a general SIC measurements for arbitrary $d$, and can be explicitly constructed \cite{GK2}.
Let $\{F_{\alpha}\}_{\alpha=1}^{d^{2}-1}$ be a set of $d^{2}-1$ Hermitian, traceless operators acting on $\mathcal{H}_{d}$,
satisfying $\mathrm{Tr}(F_{\alpha}F_{\beta})=\delta_{\alpha,\beta}$.
Define $F=\sum_{\alpha=1}^{d^{2}-1}F_{\alpha}$, then the $d^{2}$ operators
\begin{equation}
\begin{split}
P_{\alpha}&=\frac{1}{d^{2}}I+t[F-d(d+1)F_{\alpha}],~~~\alpha=1,2,\ldots,d^{2}-1,\\
P_{d^{2}}&=\frac{1}{d^{2}}I+t(d+1)F,
\end{split}
\end{equation}
form a general SIC measurements, where $t$ should be chosen such that all $P_{\alpha}\geq0$, and the parameter $a$ is given by
\begin{equation}
a=\frac{1}{d^{3}}+t^{2}(d-1)(d+1)^{3}.
\end{equation}
In the next, we characterize the BZ invariant information in general SIC measurements.

We define the quantity $V(\mathcal{P}_{\mathrm{GSM}}|\rho)$ as a measure of total variance in a general SIC measurements
$\mathcal{P}_{\mathrm{GSM}}=\{P_{\alpha}\}_{\alpha=1}^{d^{2}}$ with the parameter $a$ as follows
\begin{equation}
V(\mathcal{P}_{\mathrm{GSM}}|\rho):=\sum_{\alpha=1}^{d^{2}}V(P_{\alpha}|\rho).
\end{equation}
By using the fact that \cite{Ras1}
\begin{equation}
\sum_{\alpha=1}^{d^{2}}\langle P_{\alpha}\rangle_{\rho}^{2}=\frac{(ad^{3}-1)\mathrm{Tr}(\rho^{2})+d(1-ad)}{d(d^{2}-1)},
\end{equation}
we obtain
\begin{equation}
V(\mathcal{P}_{\mathrm{GSM}}|\rho)=\sum_{\alpha=1}^{d^{2}}\langle P_{\alpha}^{2}\rangle_{\rho}-\frac{(ad^{3}-1)\mathrm{Tr}(\rho^{2})+d(1-ad)}{d(d^{2}-1)}.
\end{equation}
On the other hand,
\begin{eqnarray*}
\sum_{\alpha=1}^{d^{2}}\langle P_{\alpha}^{2}\rangle_{\rho} & = &
\sum_{\alpha=1}^{d^{2}-1}\left\langle\left[\frac{1}{d^{2}}I+t(F-d(d+1)F_{\alpha})\right]^{2}\right\rangle_{\rho}
+\left\langle\left[\frac{1}{d^{2}}I+t(d+1)F\right]^{2}\right\rangle_{\rho}\\
& = & \frac{1}{d^{2}}+t^{2}\sum_{\alpha=1}^{d^{2}-1}\langle(F-d(d+1)F_{\alpha})^{2}\rangle_{\rho}+t^{2}(d+1)^{2}\langle F^{2}\rangle_{\rho}\\
& = & \frac{1}{d^{2}}+t^{2}d^{2}(d+1)^{2}\sum_{\alpha=1}^{d^{2}-1}\langle F_{\alpha}^{2}\rangle_{\rho}\\
& = & \frac{1}{d^{2}}+t^{2}d^{2}(d+1)^{2}(d-\frac{1}{d})\\
& = & ad.
\end{eqnarray*}
Again, in the last equality, we have used the fact that $\sum_{\alpha=1}^{d^{2}-1}F_{\alpha}^{2}=(d-\frac{1}{d})I$ \cite{Luo}.
Hence we have
\begin{equation}
V(\mathcal{P}_{\mathrm{GSM}}|\rho)=\frac{ad^{3}-1}{d(d^{2}-1)}(d-\mathrm{Tr}(\rho^{2})).
\end{equation}

The quantity $V(\mathcal{P}_{\mathrm{GSM}}|\rho)$ also enables us to characterize the BZ invariant information $I(\mathcal{P}_{\mathrm{GSM}}|\rho)$ and the BZ invariant uncertainty $U(\mathcal{P}_{\mathrm{GSM}}|\rho)$.
Taking into account that
\begin{equation}
V_{\mathrm{max}}(\mathcal{P}_{\mathrm{GSM}}|\rho)=\frac{ad^{3}-1}{d(d^{2}-1)}(d-\frac{1}{d}),~~~
V_{\mathrm{min}}(\mathcal{P}_{\mathrm{GSM}}|\rho)=\frac{ad^{3}-1}{d(d^{2}-1)}(d-1),
\end{equation}
we have
\begin{equation}
I(\mathcal{P}_{\mathrm{GSM}}|\rho)=V_{\mathrm{max}}(\mathcal{P}_{\mathrm{GSM}}|\rho)-V(\mathcal{P}_{\mathrm{GSM}}|\rho)=
\frac{ad^{3}-1}{d(d^{2}-1)}[\mathrm{Tr}(\rho^{2})-\frac{1}{d}],
\end{equation}
which is also coincide with the illustration of the BZ invariant information in general SIC measurements given in \cite{Ras}. Moreover, we have
\begin{equation}
U(\mathcal{P}_{\mathrm{GSM}}|\rho)=V(\mathcal{P}_{\mathrm{GSM}}|\rho)-V_{\mathrm{min}}(\mathcal{P}_{\mathrm{GSM}}|\rho)=
\frac{ad^{3}-1}{d(d^{2}-1)}[1-\mathrm{Tr}(\rho^{2})].
\end{equation}

\section{Conclusion}

We have studied the Brukner-Zeilinger invariant information (uncertainty) with respect to MUMs and general SIC measurements. By summing the variances over a complete set of mutually unbiased measurements and general symmetric informationally complete measurements respectively,
we have defined a new measure of information (uncertainty) in quantum measurements,
and reinterpreted the invariant information in an alternative perspective.
These results can also be used to provide an operational link between the previous interpretations of the Brukner-Zeilinger invariant information.
Taking into consideration that our new quantity of information content in measurements is defined in a simple and intuitive way, and is invariant under unitary transformations of the quantum states,
one may expect that it brings significant applications in quantum information theory.

\vspace{2.5ex}
\noindent{\bf Acknowledgments}\, \,
This work is supported by the National Natural Science Foundation of China under Grant Nos. 11805143 and 11675113, Beijing Municipal Commission of Education under Grant No. KZ201810028042.

\end{document}